\documentclass[letter,
               keeplastbox,   
               ]{jacow}

\usepackage{epsfig}
\usepackage{graphics}\graphicspath{{graph/}{}}
\usepackage{amssymb,amsmath,amsfonts}
\usepackage{time}
\usepackage{epstopdf}
\usepackage{color}
\usepackage{hyperref}
\hypersetup{
  colorlinks,
  citecolor=magenta,
  linkcolor=blue}
\usepackage{fancyhdr}
\usepackage{datetime}

\newcommand{\p}{\partial}
\newcommand{\const}{{\rm const}}

\renewcommand{\Re}{{\rm Re\,}}

\renewcommand{\Re}{{\rm Re}}

\begin{document}

\title{Suppression of the Fast Beam-Ion Instability by Tune Spread in the Electron Beam due to Beam-Beam Effects}
\author{Gennady Stupakov\thanks{stupakov@slac.stanford.edu}, SLAC National Accelerator Laboratory, Menlo Park, CA, USA}
\maketitle

\begin{abstract}
The fast beam-ion instability (FII) is caused by the interaction of an electron bunch train with the residual gas ions. The ion oscillations in the potential well of the electron beam have an inherent frequency spread due to the nonlinear profile of the potential. However, this frequency spread and associated with it Landau damping typically is not strong enough to suppress the instability. In this work, we develop a model of FII which takes into account the frequency spread in the electron beam due to the beam-beam interaction in an electron-ion collider. We show that with a large enough beam-beam parameter the fast ion instability can be suppressed. We estimate the strength of this effect for the parameters of the eRHIC electron-ion collider.

\end{abstract}

%
\section{INTRODUCTION}
%

A fast beam-ion instability (FII) which is caused by the interaction of a single electron bunch train with the residual gas ions, has been proposed and studied theoretically in Refs.~\cite{raubenheimer95z,stupakov95rz}. The instability mechanism is the same in linacs and storage rings assuming that the ions are cleared in one turn. The ions generated by the head of the bunch train oscillate in the transverse direction and resonantly interact with the betatron oscillations of the subsequent bunches, causing the growth of an initial perturbation of the beam.

An important element that has to be included into the treatment of the instability is the frequency spread in the ion population due to the nonlinearity of the potential well for the trapped ions~\cite{stupakov95rz}, as well as the spatial variation of the ion frequency along the beam path~\cite{stupakov98_2}.  This frequency spread introduces the mechanism of Landau damping, or decoherence, but does not completely suppress the instability --- it only makes it somewhat slower. In this work, we study another source of the decoherence in the fast ion instability originating from the tune spread in the electron beam. Such a tune spread may be due to the beam-beam collisions in a lepton or electron-ion collider.

For an analytical study we adopt a model that treats the bunch train as a continuous beam. This model is applicable if the distance between the bunches $l_b$ is smaller than the betatron wavelength, $l_b\ll c/\omega_\beta$, and is also smaller than the ion oscillation wavelength, $l_b\ll c/\omega_i$.  We assume a one-dimensional model that treats only vertical linear oscillation of the centroids  of the beam and the ions. We treat the electron tune spread using the method developed in Ref.~\cite{stupakov93c}.

%
\section{EQUATIONS OF MOTION}
%

We use the notation $\tilde y_e\left( {s,t|\omega_\beta} \right)$ for the vertical offset of an electron in the beam that is characterized by the betatron frequency $\omega_\beta$, at time $t$ and longitudinal position $s$. The distance $s$ is measured from the injection point at $t=0$. The equation for $\tilde y_e$, including the interaction with the ion background, is derived in Refs.~\cite{raubenheimer95z,stupakov95rz},
    \begin{align}\label{eq:1}
    &\left( {{1 \over c} {\partial  \over {\partial t}}
    +
    {\partial
    \over {\partial s}}} \right)^2
    \tilde y_e\left( {s,t|\omega_\beta} \right)
    +
    {\omega _\beta ^2 \over c^2} 
    \tilde y_e\left( {s,t|\omega_\beta} \right)
    \nonumber\\&
    =
    (ct-s)\kappa 
    \left[
    \bar y_i( s,t)
    -
    \bar y_b(s,t) 
    \right]
    .
    \end{align}
The left-hand side of this equation accounts for the free betatron oscillations of a moving beam (we assume $\varv_\mathrm{beam}\approx c$). On the right hand side, we included the force acting on the beam from the ions whose centroid is offset by $\bar y_i(s,t)$. The centroid of the electron beam is denoted by $\bar y_b(s,t)$. In the linear theory, which is the subject of this work, the interaction force between the electron beam and ions is proportional to  the relative displacement between the beam and ions centroids; it is also proportional to the ion density. Assuming a continuous electron beam with a uniform density per unit length, the ion density increases due to collisional ionization as $ct-s$ behind the head of the beam (it is equal to zero before the beam head arrives at the point $s$ at time $t=s/c$). After separating the factor $ct-s$ on the right hand side of Eq.~\eqref{eq:1}, the coefficient $\kappa $ is 
    \begin{equation}\label{eq:3}
    \kappa 
    \equiv 
    \frac{4\dot \lambda_{ion}r_e}
    {3\gamma c\sigma_y(\sigma_x+\sigma_y)}
    ,
    \end{equation}
where $\gamma$ is the relativistic factor of the beam, $r_e$  is the classical electron radius, $\sigma _{x,y}$  denote the horizontal and vertical rms-beam size respectively, and $\dot \lambda _{ion}$  is the number of ions per meter generated by the beam per unit time. Assuming a cross section for collisional ionization of about 2 Mbarns (corresponding to carbon monoxide and the electron energy $\sim 10$ GeV), we have 
    \begin{equation}\label{eq:4}
    \dot \lambda _{ion}[{\rm m^{-1}s^{-1}}]
    \approx 
    1.8\cdot 10^9
    n_e
    [{\rm m^{-1}}]
    p_{gas}
    [{\rm torr}]\ ,
    \end{equation}
where $n_e$  is the number of electrons in the beam per meter, and $p_{gas}$  the residual gas pressure in torr.

We assume that there is a betatron frequency spread in the electron beam due to the beam-beam interaction which is described by the distribution function $f_e(\omega_\beta)$ normalized by unity, $\int f_e(\omega_\beta) d\omega_\beta = 1$. The betatron frequency spread is assumed small, so that the function $ f_e(\omega_\beta)$ is localized around the central frequency $\omega_{\beta 0}$. The centroid offset is obtained through averaging $\tilde y_e( s,t|\omega_\beta)$ with the help of the distribution function,
    \begin{align}\label{eq:2}
    \bar  y_b(s,t)
    =
    \int f_e(\omega_\beta) 
    \tilde y_e( s,t|\omega_\beta)
    d\omega_\beta
    .
    \end{align}

To find the equation for ions, we will assume that they perform linear
oscillations inside the beam with a frequency $\omega _i$. Furthermore,
we will allow a continuous spectrum of $\omega _i$  given by a distribution function $f_i \left( {\omega _i} \right)$  normalized so that $\int f_i\left( {\omega _i} \right)d\omega _i=1$.
The distribution $f_i\left( {\omega _i} \right)$  is peaked around the frequency $\omega _i=\omega _{i0}$  corresponding to small vertical oscillations on the axis, 
    \begin{equation}\label{eq:5}
    \omega _{i0}
    \equiv 
    \left[ 
    \frac{4n_er_p c^2}  
    {3A\sigma _y(\sigma_x+\sigma_y)}
    \right]^{1/2}
    ,
    \end{equation}
where $A$ designates the atomic mass number of the ions, $n_e$ the number of electrons in the beam per unit length, and $r_p$  the classical proton radius ($r_p\approx 1.5\cdot 10^{-16}$ cm). Typically, the frequency spread $\Delta \omega _i$ is not large; we assume $\Delta \omega_i \ll \omega_{i0}$.

We have to distinguish between the ions generated at different times $t'$ because they will have an initial offset equal to the beam coordinate $\bar y_b\left( {s,t'} \right)$. Let us denote by $\tilde y_i\left( {s,t|t',\omega _i} \right)$ the displacement, at time $t$ and position $s$, of the ions generated at $t'$ ($t'\le t$) and oscillating with the frequency $\omega _i$. We have an oscillator equation for $\tilde y_i$
    \begin{equation}\label{eq:6}
    \frac{\partial ^2}  {\partial t^2}
    \tilde y_i(s,t|t',\omega_i)
    +
    \omega _i^2
    \left[
    {\tilde y_i(s,t|t',\omega_i)
    -
    \bar y_b(s,t)} 
    \right]
    =
    0
    ,
    \end{equation}
with the initial condition
    \begin{equation}\label{eq:7}
    \tilde y_i( {s,t'|t',\omega _i} )
    =
    \bar y_b( {s,t'}),\qquad
    \left. 
    \frac{\partial \tilde y_i}
    {\partial t}
    \right|_{t=t'}
    =
    0
    .
    \end{equation}
Finally, averaging the displacement of the ions produced at different times $t'$ and having different frequencies $\omega _i$  gives the ion centroid $\bar y_i\left( {s,t} \right)$,
    \begin{equation}\label{eq:8}
    \bar y_i( {s,t} )
    =
    \frac{1} {t-s/c}
    \int_{s/c}^t {dt'}
    \int d\omega_i 
    f_i(\omega _i)
    \tilde y_i(s,t|t',\omega_i)
    .
    \end{equation}

Equations~\eqref{eq:1},~\eqref{eq:6}-\eqref{eq:8} constitute a full set of equations governing the
development of the instability.

%
\section{AVERAGING EQUATIONS}
%

Equation~\eqref{eq:6} can be easily integrated with the initial conditions~\eqref{eq:7}
yielding
    \begin{equation}\label{eq:9}
    \tilde y_i\left( {s,t|t',\omega _i} \right)
    =
    \bar y_b\left( {s,t} \right)
    -
    \int\limits_{t'}^t 
    \frac
    {\partial \bar y_b(s,t'')}  
    {\partial t''}
    \cos \omega_i (t-t'')dt''
    .
    \end{equation}
Now using Eq.~\eqref{eq:8} and~\eqref{eq:9} in Eq.~\eqref{eq:1} we find an integro-differential equation for $\tilde y_e$,
    \begin{align}\label{eq:10}
    &\left( {{1 \over c} {\partial  \over {\partial t}}
    +
    {\partial
    \over {\partial s}}} \right)^2
    \tilde y_e\left( {s,t|\omega_\beta} \right)
    +
    {\omega _\beta ^2 \over c^2} 
    \tilde y_e\left( {s,t|\omega_\beta} \right)
    \nonumber\\&
    =
    -\kappa
    \int_{s/c}^t 
    (ct'-s)
    \frac{\partial \bar y_b(s,t')} {\partial t'}
    D_i(t-t')dt'
    ,
    \end{align}
where $D_i(t-t')$  denotes the ion decoherence function defined as
    \begin{equation}\label{eq:11}
    D_i(t-t')
    =
    \int 
    d\omega_i\cos \omega_i(t-t' )
    f_i( {\omega_i} )
    .
    \end{equation}
This function represents the oscillation of the centroid of an ensemble of ions with a given frequency distribution $f_i( {\omega _i} )$ having an initial unit offset. If there are no frequency spread in the beam ($f_i = \delta(\omega_i-\omega_{i0})$) we have $D_i(t)=\cos ( \omega_{i0} t )$.

Instead of $t$ and $s$, it is convenient to transform to new independent variables $z$ and $s$, where $z=ct-s$. The variable $z$ measures the distance from the head of the beam train and for a fixed $z$ the variable $s$ plays a role of time. Denoting 
    \begin{align}\label{eq:12}
    &y\left( {s,z} \right)
    \equiv 
    \bar y_b\left(s,\frac{1}{c}(s+z) \right)
    ,
    \nonumber\\&
    y_e\left( {s,z|\omega_\beta} \right)
    \equiv
    \tilde y_e
    \left( s,\frac{1}{c}(s+z)|\omega_\beta \right)
    ,
    \end{align}
Eq.~\eqref{eq:10} takes the form
    \begin{align}\label{eq:13}
    &{{\partial ^2} \over {\partial s^2}}
    y_e\left( {s,z} |\omega_\beta\right)
    +
    {\omega _\beta^2 \over c^2} 
    y_e\left( {s,z}|\omega_\beta \right)
    \nonumber\\&
    =
    -\kappa \int_0^z 
    {z'{{\partial y\left( {s,z'}
    \right)} \over {\partial z'}}}
    D_i[(z-z')/c]dz'
    .
    \end{align}

We will assume that the parameter $\kappa$ that defines the interaction between the beam and the ions is small,
    \begin{equation}\label{eq:14}
    c^2 \kappa l\ll\omega _{i0}^2,
    \,\,\omega _\beta ^2
    \,,
    \end{equation}
where $l$ denotes the length of the bunch train. This inequality means that the instability develops on a time scale that is much larger than both the betatron period and the period of ion oscillations. Typically this inequality is easily satisfied. In such a situation, the most unstable solution of Eq.~\eqref{eq:13} can be represented as a wave propagating in the beam with a slowly varying amplitude and phase,
    \begin{align}\label{eq:15}
    y_e(s,z|\omega_\beta)
    &=
    \Re A_e(s,z|\omega_\beta)
    e^{-i\omega _{\beta 0} s/c+i\omega_{i0}z/c}
    ,
    \end{align}
where the complex amplitude $A_e(s,z|\omega_\beta)$ is a `slow' function of its variables,
    \begin{equation}\label{eq:16}
    \left| {{{\partial \ln A_e} \over {\partial s}}} \right|
    \ll
    {\omega_{\beta 0} \over c}
    ,\  \;
    \left| {{{\partial \ln A_e} \over {\partial z}}} \right|
    \ll
    {\omega_{i0} \over c}
    .
    \end{equation}
For a fixed $z$, the $s$-dependence of Eq.~\eqref{eq:15} describes a pure betatron oscillation, while, for a fixed $s$ (that is in the frame co-moving with the ionspendent part implies oscillations (in time $t$) with the frequency $\omega _{i0}$. Hence the wave resonantly couples the oscillations of ions  and electrons. Note that from Eq.~\eqref{eq:2} it follows that for the average offset of the electron beam we have
    \begin{align}\label{eq:17}
    y(s,z)
    &=
    \Re A(s,z)
    e^{-i\omega _{\beta 0} s/c+i\omega_{i0}z/c}
    ,
    \end{align}
with
    \begin{align}\label{eq:18}
    A(s,z)
    =
    \int f_e(\omega_\beta) 
    A_e(s,z|\omega_\beta)
    d\omega_\beta
    .
    \end{align}

Substituting Eq.~\eqref{eq:15} into Eq.~\eqref{eq:13} and averaging it over the rapid oscillations
with the frequencies $\omega _{i0}$ and $\omega _{\beta 0} $, we can re-formulate Eq.~\eqref{eq:13} so that it describes a slow evolution of the complex function $A_e$,
    \begin{align}\label{eq:19}
    &
    \frac{\partial A_e(s,z|\omega_\beta)}  {\partial s}
    +
    \frac{i}{c}
    (\omega_\beta-\omega_{\beta 0})
    A_e(s,z|\omega_\beta)
    \nonumber\\&
    =
    \frac{\kappa \omega_{i0}}  
    {4\omega_{\beta 0} }
    \int_0^z 
    z'A(s,z')
    \hat D_i(z-z')dz'
    ,
    \end{align}
where the function $\hat D_i( z)$ is
    \begin{equation}\label{eq:20}
    \hat D_i\left( z \right)
    =
    \int {d\omega _i
    f_i( {\omega _i} )
    e^{i(\omega _i-\omega _{i0})z/c}}
    .
    \end{equation}
Eqs.~\eqref{eq:18} and~\eqref{eq:19} constitute a full set of equations that we need to solve.

We can make one more step and formulate an equation for the amplitude of the averaged offset $A$. For this, we integrate Eq.~\eqref{eq:19} over $s$,
    \begin{align}\label{eq:21}
    &A_e(s,z|\omega_\beta)
    =
    A_e(0,z|\omega_\beta)
    e^{-i(\omega_\beta-\omega_{\beta 0})s/c}
    \nonumber\\&
    +
    \frac{\kappa \omega_{i0}}  
    {4\omega_{\beta 0} }
    \int_{0}^{s}
    ds'
    e^{-i(\omega_\beta-\omega_{\beta 0})(s-s')/c}
    \nonumber\\&\times
    \int_0^z 
    z'A(s',z')
    \hat D_i(z-z')dz'
    .
    \end{align}
We now average this equation with the distribution function $f_e(\omega_\beta)$. It is reasonable to assume that the initial offset $A_e(0,z|\omega_\beta)$ does not depend on $\omega_\beta$, and write it as $A_0(z)$. We then obtain
    \begin{align}\label{eq:22}
    &A(s,z)
    =
    A_0(z)
    \hat D_e(s)
    +
    \frac{\kappa \omega_{i0}}  
    {4\omega_{\beta 0} }
    \int_{0}^{s}
    ds'
    \hat D_e(s-s')
    \nonumber\\&\times
    \int_0^z 
    z'A(s',z')
    \hat D_i(z-z')dz'
    ,
    \end{align}
where
    \begin{align}\label{eq:23}
    \hat D_e(s)
    =
    \int d\omega_\beta
    f_e(\omega_\beta)
    e^{-i(\omega_\beta-\omega_{\beta 0})s/c}
    .
    \end{align}
Note that from Eq.~\eqref{eq:22} follows the initial condition for function $A$,
    \begin{align}\label{eq:24}
    &A(s,0)
    =
    A_0(0)
    \hat D_e(s)
    .
    \end{align}

We can also re-write Eq.~\eqref{eq:22} as an integro-differential equation
    \begin{align}\label{eq:25}
    &\frac{\p}{\p s}
    A(s,z)
    =
    A_0(z)
    \hat D_e'(s)
    +
    \frac{\kappa \omega_{i0}}  
    {4\omega_{\beta 0} }
    \int_0^z 
    z'A(s,z')
    \hat D_i(z-z')dz'
    \nonumber\\&
    +
    \frac{\kappa \omega_{i0}}  
    {4\omega_{\beta 0} }
    \int_{0}^{s}
    ds'
    \hat D_e'(s-s')
    \int_0^z 
    z'A(s',z')
    \hat D_i(z-z')dz'
    ,
    \end{align}
where the prime denotes the derivative with respect to $s$, and we have used $\hat D_e(0) = 1$. The first and the third terms on the right-hand side vanish for a constant $\hat D_e$ that corresponds to the case of the zero electron tune spread, and in this limit we recover the result of Ref.~\cite{stupakov95rz}.

%
\section{SOLUTION OF FII EQUATIONS for special cases}
%

In this section will show how to solve Eq.~\eqref{eq:22} for the case when one can neglect the ion decoherence, $\hat D_i(z)=1$. In this case Eq.~\eqref{eq:22} reduces to
    \begin{align}\label{eq:26}
    &A(s,z)
    =
    A_0(z)
    \hat D_e(s)
    +
    \frac{\kappa \omega_{i0}}  
    {4\omega_{\beta 0} }
    \nonumber\\&\times
    \int_{0}^{s}
    ds'
    \hat D_e(s-s')
    \int_0^z 
    z'A(s',z')dz'
    .
    \end{align}
We first make the Laplace transform with respect to the variable $s$, introducing the Laplace image $a(\varkappa,z)$,
    \begin{align}\label{eq:27}
    a(\varkappa,z)
    =
    \int_{0}^{\infty}
    A(s,z)
    e^{-\varkappa s}
    ds.
    \end{align}
Making the Laplace transform of Eq.~\eqref{eq:26} we find
    \begin{align}\label{eq:28}
    a(\varkappa,z)
    =
    A_0(z)
    d(\varkappa)
    +
    \frac{\kappa \omega_{i0}}  
    {4\omega_{\beta 0} }
    d(\varkappa)
    \int_0^z 
    z'a(\varkappa,z')dz'
    ,
    \end{align}
where 
    \begin{align}\label{eq:29}
    d(\varkappa)
    =
    \int_{0}^{\infty}
    \hat D_e(s)
    e^{-\varkappa s}
    ds.
    \end{align}

We can solve Eq.~\eqref{eq:28} analytically for the special case when the initial amplitude of the beam offset, $A_0$, does not depend on $z$, $A_0=\const$. Differentiating Eq.~\eqref{eq:28} with respect to $z$ and solving the resulting differential equation with the initial condition $a(\varkappa,0) = A_0 d(\varkappa)$ gives the following result
    \begin{align}\label{eq:30}
    a(\varkappa,z)
    =
    A_0
    d(\varkappa)
    e^{qd(\varkappa)z^2}
    ,
    \end{align}
with $q ={\kappa \omega_{i0}}/{8\omega_{\beta 0} }$. Making the inverse Laplace transform, we can find $A$,
    \begin{align}\label{eq:31}
    A(s,z)
    =
    A_0
    \frac{1}{2\pi i}
    \int_{\sigma-i\infty}^{\sigma+i\infty}
    d\varkappa
    e^{\varkappa s}
    d(\varkappa)
    e^{qd(\varkappa)z^2}
    .
    \end{align}

In the case when the tune spread in the electron beam is so small that it can be neglected, we have $\hat D_e(s) = 1$ and $d(\varkappa) = 1/\varkappa$. We arrive at the integral
    \begin{align}\label{eq:32}
    \frac{1}{2\pi i}
    \int_{\sigma-i\infty}^{\sigma+i\infty}
    \frac{d\varkappa}{\varkappa}
    e^{\varkappa s+qz^2/\varkappa}
    =
    I_0
    \left(
    2z
    \sqrt{{q}{s}}
    \right)
    ,
    \end{align}
and $A=A_0I_0  \left( 2z \sqrt{{q}{s}}  \right)$. This is the result of Refs.~\cite{raubenheimer95z, stupakov95rz} when the ion frequency spread is neglected.

Consider now a model electron decoherence function
    $
    \hat D_e(s) 
    =
    e^{-ps}
    $
with $p>0$, for which we have $d(\varkappa) = 1/(\varkappa + p)$. For the integral we have
    \begin{align}\label{eq:34}
    A(s,z)
    &=
    \frac{A_0}{2\pi i}
    \int_{\sigma-i\infty}^{\sigma+i\infty}
    \frac{d\varkappa}{\varkappa+p}
    e^{\varkappa s+qz^2/(\varkappa+p)}
    \\\nonumber&
    =
    A_0
    e^{-ps}
    I_0
    \left(
    2z
    \sqrt{{q}{s}}
    \right)
    .
    \end{align}
Asymptotically, in the limit $s\to\infty$, we have $I_0 ( 2z \sqrt{{q}{s}})\propto \exp( 2z \sqrt{{q}{s}})/\sqrt{ 2z \sqrt{{q}{s}}}$, and the exponential factor $e^{-ps}$ overcomes the growing Bessel function, and hence, suppresses the instability.

%
\section{Decoherence due to beam-beam collisions}
%

\begin{figure}[htb!]
	\centering
	\includegraphics[width=0.4\textwidth, trim=0mm 0mm 0mm 0mm, clip]{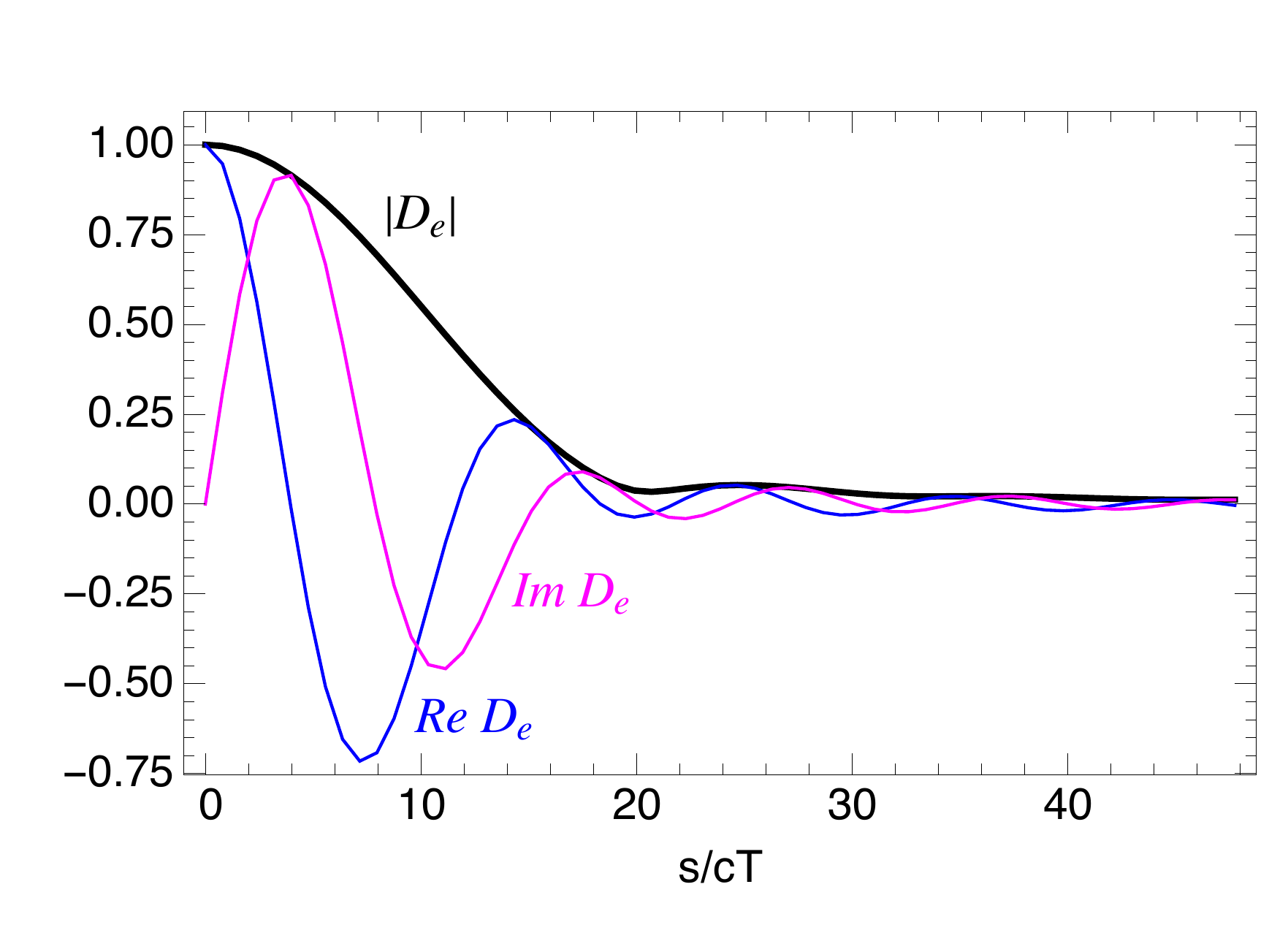}
	\caption{Plot of the real (black) and imaginary (magenta) parts of the function $\hat{D}_e(s)$. The black line is the absolute value $|\hat{D}_e(s)|$.}
	\label{fig:1}
	\end{figure}

The decoherence function $\hat{D}_e(s)$ for the case when the betatron tune spread is due to the beam-beam collisions at the interaction point in a collider was derived in Ref.~\cite{stupakov93ps}. Assuming round beams at the interaction point, the following expression for $\hat{D}_e(s)$ was obtained:
    \begin{align}\label{eq:35}
    \hat{D}_e(s)
    &=
    4
    \int_{0}^{\infty}
    \int_{0}^{\infty}
    da_1\,da_2
    \nonumber\\&\times
    \exp
    \left[
    -2(a_1+a_2)
    +
    i(s/cT)
    \Delta\nu_y(a_1,a_2)
    \right]
    ,
    \end{align}
where $T$ is the revolution period in the ring and the tune shift $\Delta\nu(a_1,a_2)$ is given by the following formula~\cite{chao83},
    \begin{align*}
    \Delta\nu
    =
    \xi
    \int_{0}^{1}
    du\,
    e^{-u(a_1+a_2)}
    I_0(a_2 u)
    \left[
    I_0(a_1 u)
    -
    I_1(a_1 u)
    \right]
    .
    \end{align*}
Here $a_1$ and $a_2$ are the dimensionless amplitudes of the betatron oscillations, $I_n(z)$ is the modified Bessel function of the $n$-th order and $\xi$ is the tune shift parameter, $\xi = N_pr_e/4\pi\epsilon$ with $N_p$ the number of particles in the bunch, $r_e$ the classical electron radius and $\epsilon$ the normalized beam emittance. The tune shift $\Delta\nu$ is positive due to the opposite signs of the charges of the colliding beams in an electron-ion collider. The plot of function $\hat{D}_e(s)$ is shown in Fig.~\ref{fig:1}.

%
\section{NUMERICAL SOLUTION OF EQ.~(\ref{eq:25})}
%

Here we outline the numerical solution of~\eqref{eq:25} following the method proposed in Ref.~\cite{numerical_sol}. We first normalize the variable $z$ by the length of the bunch train $l$, $\zeta = z/l$, and then replace $s$ in this equation by $\xi = s({\kappa \omega_{i0}l^2}/    {4\omega_{\beta 0} })$. Note that positions within the bunch train correspond to the interval $0<\zeta<1$. We then have
    \begin{align}\label{eq:A.1}
    &\frac{\p}{\p \xi}
    A(\xi,\zeta)
    =
    A_0(\zeta)
    \hat D_e'(\xi)
    +
    \int_0^\zeta 
    \zeta'A(\xi,\zeta')
    \hat D_i(\zeta-\zeta')d\zeta'
    \nonumber\\&
    +
    \int_{0}^{\xi}
    d\xi'
    \hat D_e'(\xi-\xi')
    \int_0^\zeta 
    \zeta'A(\xi',\zeta')
    \hat D_i(\zeta-\zeta')d\zeta'
    ,
    \end{align}
where $\hat D_e'$ now denotes the derivative of $\hat D_e$ with respect to $\xi$. 

We first introduce a mesh in the unit interval $0<\zeta<1$, $\zeta_k = k\Delta$, $k=1,\ldots,n$, with $\Delta = 1/(n-1)$. The function $A(\xi,\zeta)$ is now represented on this mesh, $A(\xi,\zeta_k)$, and we use the trapezoidal integration rule to carry out the integration over $\zeta$ in Eq.~\eqref{eq:25}, 
    \begin{align}\label{eq:A.2}
    &\int_0^\zeta 
    \zeta'A(\xi',\zeta')
    \hat D_i(\zeta-\zeta')d\zeta'
    \nonumber\\
    &=
    \frac{\Delta}{2}
    \left[
    \zeta_1A(\xi',\zeta_1)
    +
    2\zeta_2A(\xi',\zeta_2)
    +
    \ldots
    \right.
    \nonumber\\&
    \left.
    +
    2\zeta_{n-1}A(\xi',\zeta_{n-1})
    +
    \zeta_{n}A(\xi',\zeta_{n})
    \right]
    .
    \end{align}
We will use the notation ${\cal A}(\xi)$ for the vector $A(\xi,\zeta_k)$ and denote the discretized integration~\eqref{eq:A.2} by the operator $T$,
    \begin{align}\label{eq:A.3}
    \int_0^\zeta 
    \zeta'A(\xi',\zeta')
    \hat D_i(\zeta-\zeta')d\zeta'
    \to
    T\cdot{\cal A}
    .
    \end{align}
Then Eq.~\eqref{eq:A.1} can be written as
    \begin{align}\label{eq:A.4}
    {\cal A}'(\xi)
    &=
    {\cal A}_0
    \hat D_e'(\xi)
    +
    T\cdot{\cal A}(\xi)
    +
    \int_{0}^{\xi}
    d\mu
    {\cal D}(\xi,\mu)
    T\cdot{\cal A}(\mu)
    ,
    \end{align}
where to simplify the notation we introduced ${\cal D}(\xi,\mu) \equiv \hat D_e'(\xi-\mu)$.

Introducing the step $h$ in variable $\xi$ we denote by ${\cal A}_k$ the value of ${\cal A}$ at $\xi_k = h(k-1)$. Likewise ${\cal A}'_k$ denotes the value of ${\p\cal A/\p\xi}|_{\xi=\xi_k}$. Again, using the trapezoidal method of integration, we find
    \begin{align}\label{eq:A.5}
    {\cal A}'_k
    &=
    {\cal A}_0
    \hat D_e'(\xi_k)
    +
    T\cdot{\cal A}_k
    +
    \frac{h}{2}
    \left[
    {\cal D}(\xi_k,\xi_1)
    T\cdot{\cal A}_1
    \right.
    \nonumber\\
    &\left.
    +
    2{\cal D}(\xi_k,\xi_2)
    T\cdot{\cal A}_2
    +
    \ldots
    +
    2{\cal D}(\xi_k,\xi_{k-1})
    T\cdot{\cal A}_{k-1}
    \right.
    \nonumber\\
    &\left.
    +
    2{\cal D}(\xi_k,\xi_k)
    T\cdot{\cal A}_k
    \right]
    .
    \end{align}

To advance the step from $\xi=\xi_k$ to $\xi = \xi_k+h$ we first integrate~\eqref{eq:A.4} to obtain
    \begin{align}\label{eq:A.6}
    {\cal A}_{k+1}
    &=
    {\cal A}_{k}
    +
    {\cal A}_0
    [\hat D_e(\xi_{k+1})-\hat D_e(\xi_{k})]
    +
    \int_{\xi_k}^{\xi_{k+1}}
    d\xi
    T\cdot{\cal A}(\xi)
    \nonumber\\
    &+
    \int_{\xi_k}^{\xi_{k+1}}
    d\xi
    \int_{0}^{\xi}
    d\mu
    {\cal D}(\xi,\mu)
    T\cdot{\cal A}(\mu)
    \nonumber\\
    &=
    {\cal A}_{k}
    +
    {\cal A}_0
    [\hat D_e(\xi_{k+1})-\hat D_e(\xi_{k})]
    +
    I_1
    +
    I_2
    .
    \end{align}
For $I_1$ we use the trapezoidal rule together with the approximation ${\cal A}_{k+1} = {\cal A}_{k} + h{\cal A}'_k$ to yield
    \begin{align}\label{eq:A.7}
    I_1
    =
    \frac{h}{2}
    \left[
    T\cdot{\cal A}_k
    +
    T\cdot({\cal A}_k + h{\cal A}'_k)
    \right]
    .
    \end{align}
For $I_2$ we use the trapezoidal rule for the outer integral to obtain
    \begin{align}\label{eq:A.8}
    I_2
    &=
    \frac{h}{2}
    \left[
    \int_{0}^{\xi_k}
    d\mu
    {\cal D}(\xi_k,\mu)
    T\cdot{\cal A}(\mu)
    \right.
    \nonumber\\
    &\left.
    +
    \int_{0}^{\xi_{k+1}}
    d\mu
    {\cal D}(\xi_{k+1},\mu)
    T\cdot{\cal A}(\mu)
    \right]
    .
    \end{align}
We then use the trapezoidal rule for the inner integral and again use the approximation ${\cal A}_{k+1} = {\cal A}_{k} + h{\cal A}'_k$ to obtain
    \begin{align}\label{eq:A.9}
    &I_2
    =
    \frac{h^2}{4}
    \left[
    {\cal D}(\xi_k,\xi_1)
    T\cdot{\cal A}_1
    +
    2{\cal D}(\xi_k,\xi_2)
    T\cdot{\cal A}_2
    \ldots
    \right.
    \nonumber\\
    &\left.
    +
    2{\cal D}(\xi_k,\xi_{k-1})
    T\cdot{\cal A}_{k-1}
    +
    {\cal D}(\xi_k,\xi_k)
    T\cdot{\cal A}_k
    \right]
    \nonumber\\
    &+
    \frac{h^2}{4}
    \left[
    {\cal D}(\xi_{k+1},\xi_1)
    T\cdot{\cal A}_1
    +
    2{\cal D}(\xi_{k+1},\xi_2)
    T\cdot{\cal A}_2
    \ldots
    \right.
    \\
    &\left.
    +
    2{\cal D}(\xi_{k+1},\xi_{k})
    T\cdot{\cal A}_{k}
    +
    {\cal D}(\xi_{k+1},\xi_{k+1})
    T\cdot({\cal A}_{k} + h{\cal A}'_k)
    \right]
    .\nonumber
    \end{align}
Eqs.~\eqref{eq:A.6}, \eqref{eq:A.7} and~\eqref{eq:A.9} finalize the one step advance in the numerical solution of Eq.~\eqref{eq:25}.

%
\section{FII at {e}RHIC}
%

In this section we will analyze the fast ion instability for the electron storage ring of the proposed Electron Ion Collider at BNL, eRHIC~\cite{Montag:IPAC2017}. The parameters of eRHIC electron beam relevant for the fast ion instability are summarized in Table~\ref{tab:1}. The nominal tune shift for the electron beam is $\xi = 0.1$.
\begin{table}[hbt]
\begin{center}
\begin{tabular}{lc}
\hline
\hline
Electron beam energy\hspace{10mm} 	   &  10 GeV \\
Vertical beam emittance, $\epsilon_y$  	   &  4.9 nm \\
Horizontal beam emittance, $\epsilon_x$  	   &  20 nm \\
Residual gas pressure, $p$ & 0.75 nTorr\\
Averaged beta function, $\beta_x$, $\beta_y$ & 18 m\\
Vertical betatron tune, $\nu_y$ & 31.06\\
Number of electron bunches, $N_b$ & 567\\
Length of the bunch train, $l_b$& 3451 m\\
Atomic mass number for ions, $A$& 28\\
Number of electrons per unit length, $n_e$& $5.6\times 10^{10}$ m$^{-1}$\\
\hline
\hline
\end{tabular}
\caption{Parameters of the eRHIC collider relevant for FII.}
\label{tab:1}
\end{center}
\end{table}

Using the beam emittance and the value for the averaged beta functions we find the characteristic beam sizes in the vertical and horizontal directions, $\sigma_y = 0.3$ mm and $\sigma_x = 0.6$ mm. From Eq.~\eqref{eq:5} we obtain the ion frequency $\omega_{i0} = 4.5\times 10^7$ s$^{-1}$ and for the vertical betatron frequency we calculate $\omega_\beta = 1.5\times 10^7\ \mathrm{s}^{-1}$. From Eq.~\eqref{eq:4} it follows that $\dot\lambda = 7.5\times 10^{10}\ \mathrm{m}^{-1}\mathrm{s}^{-1}$ and  the characteristic time of FII\cite{stupakov95rz} is
    \begin{align}
    \tau
    =
    \frac{4\omega_{\beta 0}}{\kappa \omega_{i0}cl^2}
    =
    2.1\ \mu\mathrm{s}
    .
    \end{align}
In our calculations we assume the worst case scenario when all residual gas pressure is due to the carbon monoxide (lighter ions are usually less trapped inside the beam). 

Note that the parameter $c/\omega_\beta \approx 20 $ m is several times larger than the distance between the electron bunches, $l_b = 6.1$ m, but $c/\omega_{i0} = 6.5$ m is comparable with $l_b$, which means that, for the parameters of eRHIC, the model of continuous electron beam that we use in this paper is actually at the edge of its applicability range.

\begin{figure}[htb!]
	\centering
	\includegraphics[width=0.4\textwidth, trim=0mm 0mm 0mm 0mm, clip]{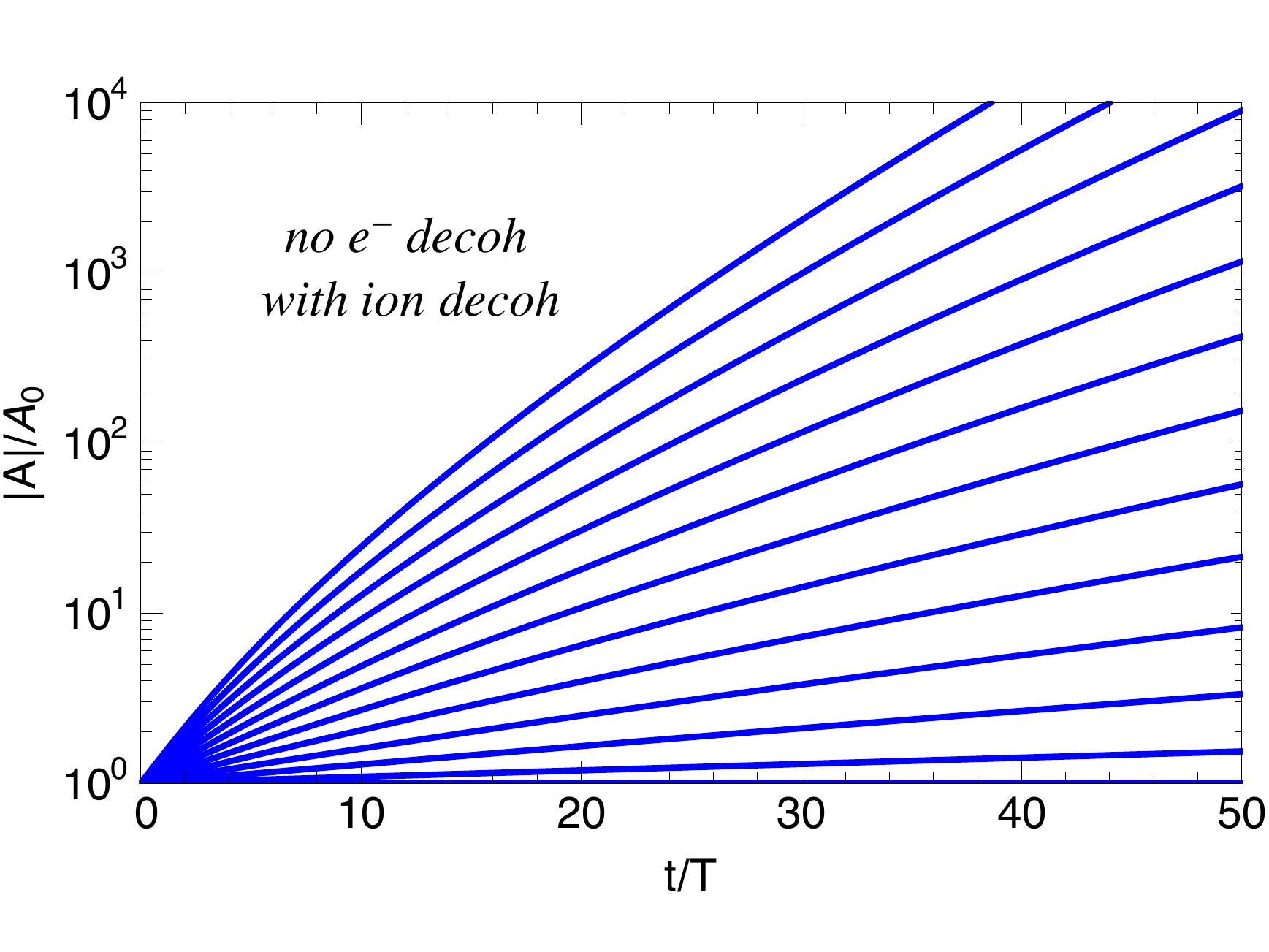}
	\caption{Amplitude $A$ normalized by its initial value $A_0$ at 13 equidistant positions in the bunch train as a function of time measured in the revolution periods $T$ in the ring. Electron decoherence effects are neglected. The amplitude grows faster with increase of the distance from the head of the train.}
	\label{fig:2}
	\end{figure}

We first simulated the fast ion instability for eRHIC parameters neglecting the electron decoherence by taking into account only ion decoherence effects. This case is described by Eq.~\eqref{eq:25} with $\hat D_e(s) = 1$. For the ion decoherence function $\hat D_i(z)$, following Ref.\cite{stupakov95rz}, we took
    \begin{align}\label{eq:45}
    \hat D_i(z)
    =
    \left(
    1
    +
    \frac{i}{4c}
    \omega_{i0}z
    \right)^{-1/2}
    .
    \end{align}
Fig.~\ref{fig:2} shows the plot of the FII amplitude obtained by numerical solution of Eq.~\eqref{eq:25} for this case at different locations along the bunch train.
This plot shows that the amplitude $A$ at the end of the electron bunch train grows more then four order of magnitude after 50 revolution periods in the ring.
 
\begin{figure}[htb!]
	\centering
	\includegraphics[width=0.4\textwidth, trim=0mm 0mm 0mm 0mm, clip]{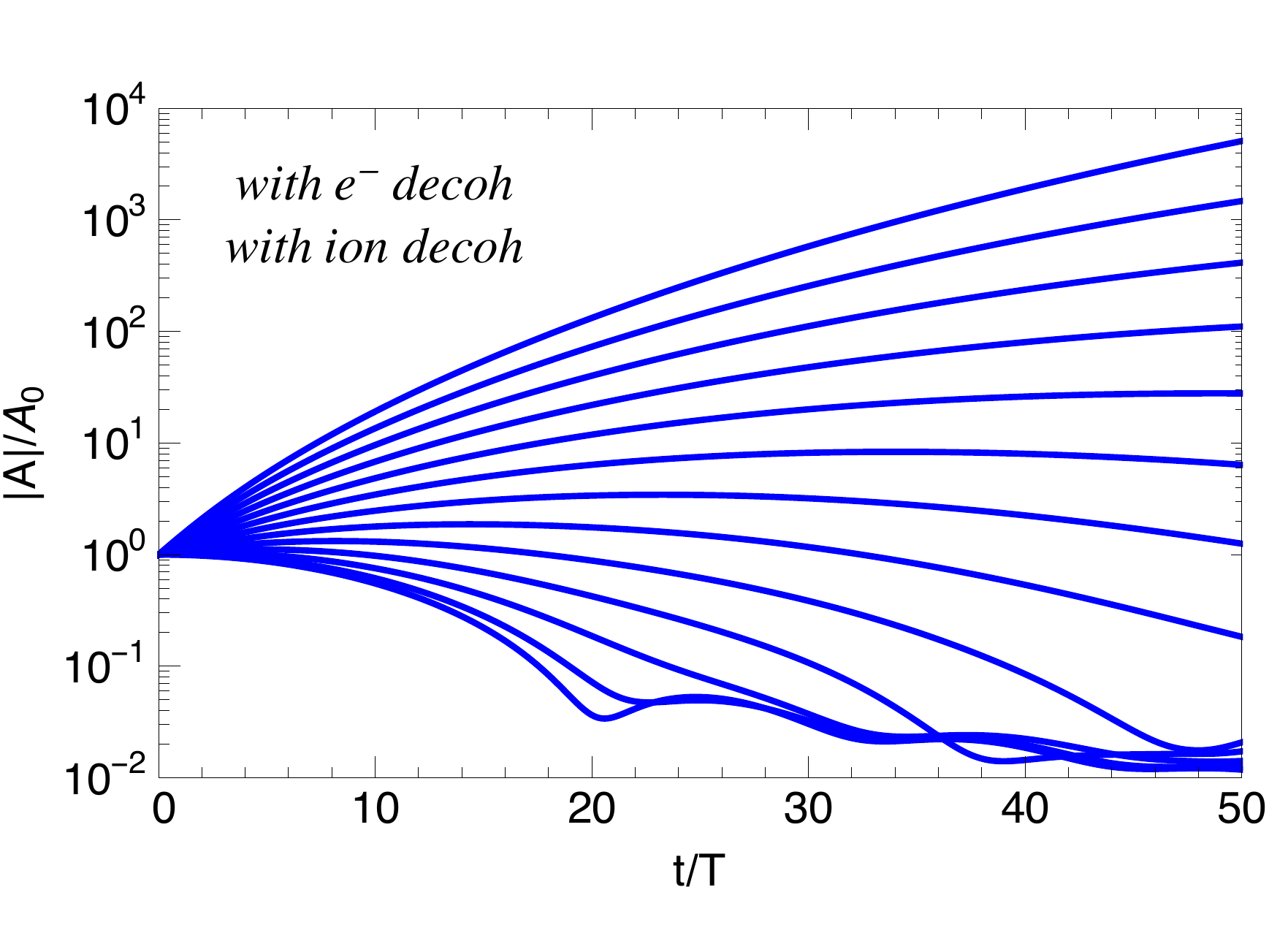}
	\caption{Amplitude $A$ normalized by its initial value $A_0$ at 13 equidistant positions in the bunch train as a function of time measured in the revolution periods $T$ in the ring.  Electron decoherence effects are taken into account.}
	\label{fig:3}
	\end{figure}

Using Fig.~\ref{fig:2} as a reference case, we then simulated FII with account of the electron beam decoherence. This case is described by Eq.~\eqref{eq:25} with the electron decoherence function~\eqref{eq:35}. For the ion decoherence function we again used Eq.~\eqref{eq:45}.  The result is shown in Fig.~\ref{fig:3}. 
One can see that the electron decoherence suppresses the instability in the region close to the head of the bunch train (the lowest 4-5 lines in the plot corresponding to positions near the head). However, the amplitude $A$ still grows to unacceptably large values at the tail of the train. Simulations show that if the numerical solution is continued to even larger values of $t$, the amplitude $A$ starts to decrease, but at its maximum at intermediate times, it is amplified by many orders of magnitude relative to the initial value. We conclude that  while electron decoherence effects do provide some stabilization effect, it is not sufficient, for the nominal parameters of Table~\ref{tab:1} (and 100\% carbon monoxide residual gas), to fully suppress FII.

\begin{figure}[htb!]
	\centering
	\includegraphics[width=0.4\textwidth, trim=0mm 0mm 0mm 0mm, clip]{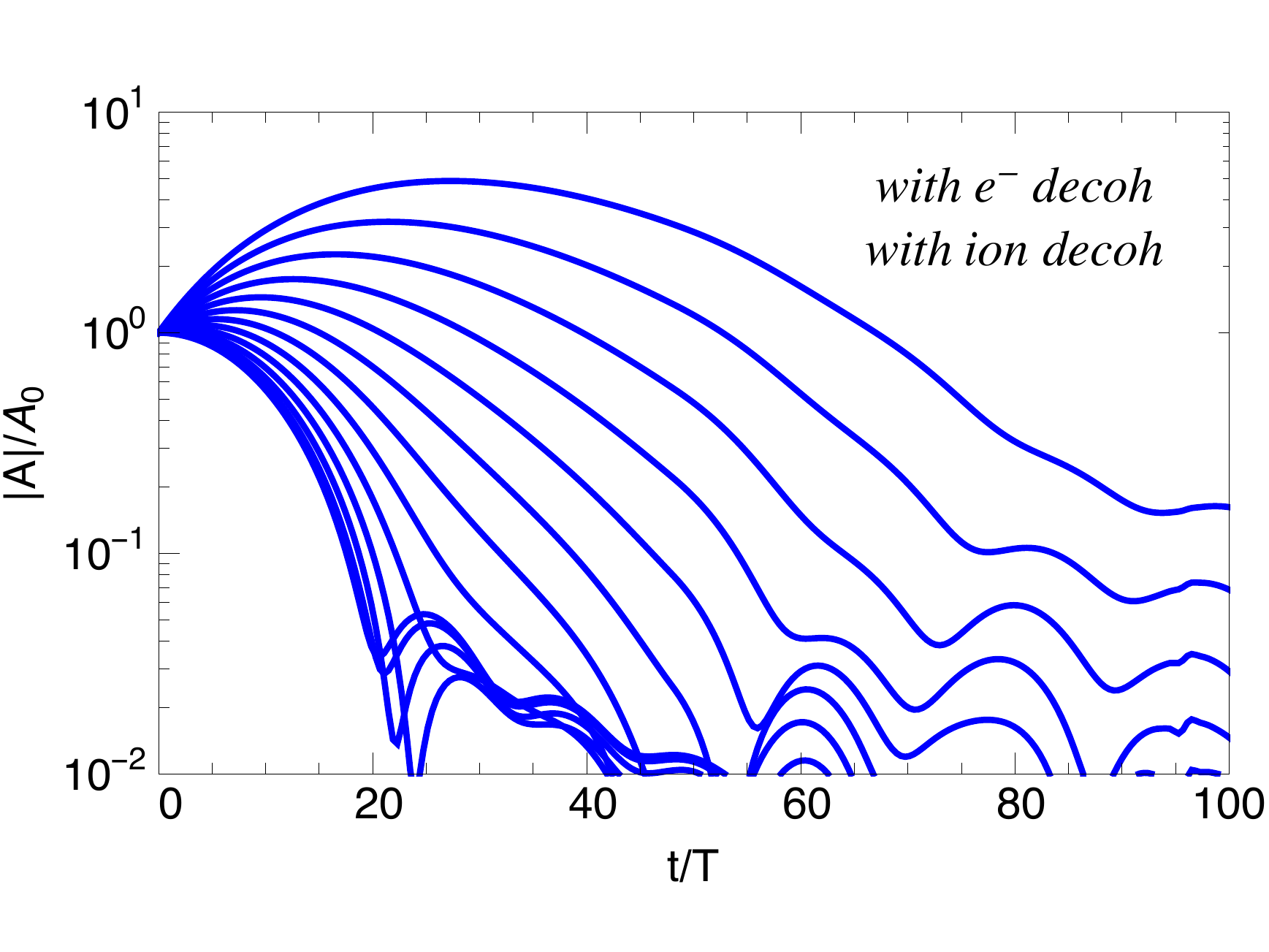}
	\caption{Amplitude $A$ normalized by its initial value $A_0$ at 13 equidistant positions in the bunch train as a function of time measured in the revolution periods $T$ in the ring.  Electron decoherence effects are taken into account. The residual gas pressure is $p = 0.25$ nTorr.}
	\label{fig:4}
	\end{figure}

Finally, we repeated the previous simulation, but with 3 times smaller residual gas pressure, $p = 0.25$ nTorr. The result is shown in Fig.~\ref{fig:4}.
This case can be characterized as a stable one: after some increase of the amplitude in the tail of the bunch train, it is suppressed, through the electron decoherence effects, to the values below the initial value $A_0$.

%
\section{SUMMARY}
%

In this paper, we extended the theoretical analysis of Ref.~\cite{stupakov95rz} of the fast ion instability to include decoherence effects due to the tune spread in the electron beam. Specifically, we calculated the electron decoherence function for the case when the tune spread is caused by the beam-beam collisions in an electron-ion collider. We derived an equation that governs the evolution of the amplitude of the transverse oscillations in the beam, and numerically solved it for the nominal parameters of the eRHIC collider. We found that while the electron decoherence weakens the instability, it does not fully suppress it for the nominal parameters of {eRHIC}. The instability however is suppressed for three times smaller residual gas pressure.
    
We note that our model of continuous electron beam is not fully applicable for eRHIC parameter because the distance between the electron bunches is of the same order as the parameter $c/\omega_{i0}$. 

Qualitatively similar results have been recently obtained by M. Blaskiewicz~\cite{Blaskiewicz:NAPAC2019} who used computer simulations of FII for eRHIC.
    
%
\section{ACKNOWLEDGMENTS}
%

The author thanks M. Blaskiewicz who initiated this work and provided relevant information for the eRHIC project, and B. Podobedov for useful discussions.

This work was supported by Department of Energy contract DE-AC03-76SF00515.

\appendix

\bibliography{\string~/gsfiles/Bibliography/master%
}

\end{document}